\documentstyle[12pt,amssymbols]{article}
\input math_macros.tex

\begin{document}
\begin{titlepage}
\begin{center}
\today     \hfill    LBL-36809 \\
%          \hfill    UCB-PTH-xx/xx \\

\vskip .5in

{\large \bf Weak Classical-Gravity Source in Standpoint Cosmology}
\footnote{This work was supported by the Director, Office of Energy
Research, Office of High Energy and Nuclear Physics, Division of High
Energy Physics of the U.S. Department of Energy under Contract
DE-AC03-76SF00098.}
\vskip .50in

%alternate footnote for faculty:
%\footnote{This work was supported in part by the Director, Office of
%Energy Research, Office of High Energy and Nuclear Physics, Division of
%High Energy Physics of the U.S. Department of Energy under Contract
%DE-AC03-76SF00098 and in part by the National Science Foundation under
%grant PHY-90-21139.}

\medskip

G.F. Chew

{\em Theoretical Physics Group\\
    Lawrence Berkeley Laboratory\\
      University of California\\
    Berkeley, California 94720}
\end{center}

\vskip .5in

\begin{abstract}
Guided by a linearized approximation to Einstein theory, an interim
prescription
for ``weak source of gravity'' - -  in ``particle'' energy-momentum distributed
along standpoint light cone - -  is formulated for (classical) standpoint
cosmology.

\end{abstract}
\end{titlepage}
%THIS PAGE (PAGE ii) CONTAINS THE LBL DISCLAIMER
%TEXT SHOULD BEGIN ON NEXT PAGE (PAGE 1)
\renewcommand{\thepage}{\roman{page}}
\setcounter{page}{2}
\mbox{ }

\vskip 1in

\begin{center}
{\bf Disclaimer}
\end{center}

\vskip .2in

\begin{scriptsize}
\begin{quotation}
This document was prepared as an account of work sponsored by the United
States Government. While this document is believed to contain correct
 information, neither the United States Government nor any agency
thereof, nor The Regents of the University of California, nor any of their
employees, makes any warranty, express or implied, or assumes any legal
liability or responsibility for the accuracy, completeness, or usefulness
of any information, apparatus, product, or process disclosed, or represents
that its use would not infringe privately owned rights.  Reference herein
to any specific commercial products process, or service by its trade name,
trademark, manufacturer, or otherwise, does not necessarily constitute or
imply its endorsement, recommendation, or favoring by the United States
Government or any agency thereof, or The Regents of the University of
California.  The views and opinions of authors expressed herein do not
necessarily state or reflect those of the United States Government or any
agency thereof, or The Regents of the University of California.
\end{quotation}
\end{scriptsize}

\vskip 2in

\begin{center}
\begin{small}
{\it Lawrence Berkeley Laboratory is an equal opportunity employer.}
\end{small}
\end{center}

\newpage
\renewcommand{\thepage}{\arabic{page}}
\setcounter{page}{1}

\noindent{\bf Introduction}

\medskip

An alternative cosmology featuring a notion of ``standpoint'' has been
proposed$^{(1)}$  and Hubble-scale predictions in ``homogeneous-universe''
approximation have been deduced.$^{(2)}$
Going beyond the homogeneous approximation requires some equivalent to the
Einstein relation between local energy-momentum tensor and metric curvature.
In lieu of a systematic deduction from quantum underpinning, we here formulate
an interim prescription - - through energy-momentum of ``particles'' - - for
``weak
local gravity source'' in classical standpoint cosmology;
for source distances small on Hubble scale, the prescription
concurs with  the linearized (weak gravity) approximation to Einstein theory.
 ``Particle'' source for the gravity experienced near standpoint is distributed
along (backward and forward) standpoint light cone.
Treatment of strong gravity will require attention to the non-Riemannian
character of standpoint metric.

In the proposed prescription the gravitational constant $G$ appears in its
traditional role of ad-hoc classical parameter.
Reference (1) indicates on the other hand that, when and if classical
standpoint
cosmology is {\it derived} from quantum underpinning, $G$ on the scale of {\it
quantum}-particle masses will relate inversely to the huge dimensionless
dilation-generator expectation that gives the ratio between big-bang scale and
the scale at which meaning began to develop for localized matter within
(classical) spacetime.
Hugeness of this scale ratio being essential in the quantum model to emergence
of spacetime, smallness of  $G$ appears essential to geometry.
Classical correction to the ``weak-gravity'' prescription proposed here is
then problematic.
In our concluding section we emphasize that, because meaning for ``particle
energy-momentum'' recognizes gravitation, ``source of gravity'' need not
exhibit any {\it explicit} gravitational ingredient {\it in addition to}
``particles''.
Nevertheless, meaning for the ``light cone'' central to our prescription must
be
reconsidered when gravitational potential is large; in its dependence on light
cone our prescription manifests a ``flat-space'' character with respect to
gravitational-signal propagation.

\newpage

\noindent{\bf II. Prescription}

\medskip

Because the wave function of standpoint quantum cosmology depends on the
energy-momentum of particles whose trajectories intersect the standpoint's
light
cone, it is natural to seek a representation of classical-gravity
source distributed along this cone.
In weak-gravity
approximation to Einstein theory when sources are confined to a
finite region of spacetime, a light-cone-source representation has been found
for the quantity

$$
 h_{\mu\nu} ({\bf{x}}) \equiv g_{\mu\nu} (\bf{x}) - \eta_{\mu\nu}\eqno(II.1)
$$
$(g_{\mu\nu} (\bf{x})$ being local metric tensor and $\eta_{\mu\nu}$ Minkowski
tensor),
through a formula of Lienard-Wiechert type.$^{(3)}$
We shall refer to the dimensionless $h_{\mu\nu} (x)$ as ``gravitational
potential''.
(Units with $c=1$ are implicit.)
In adapting to standpoint cosmology, we consider metric {\it in} ${\bf{R}}$
{\it spacetime} at a standpoint labeled ${\bf{R}}$ and exploit the model
feature
that, in {\it homogeneous-universe} approximation, a certain ``metric tensor''
denoted $g_{\mu\nu}(\bf{R})$ is equal to $\eta_{\mu\nu}$.$^{(1)}$
Our prescription is correspondingly such that $h_{\mu\nu}(\bf{R})$, defined by
an analogue of (II.1), is generated by {\it inhomogeneity} of matter
distribution.

Use of an integral formula for source rather than a differential equation is
natural because the spacetime belonging to a standpoint is finite (compact)
with
most matter concentrated near the boundary.$^{(1,2)}$
In Mach spirit one may think of this latter huge and maximally-distant source
for metric near standpoint as ``chiefly responsible'' for the Minkowski
component $\eta_{\mu\nu}$.
More precisely, in building the gravitational potential at standpoint
$\bf{R}$ (in $\bf{R}$ spacetime), defined by
$$
h_{\mu\nu}(\bf{
R}) \equiv g_{\mu\nu} (\bf{
R}) - \eta_{\mu\nu},\eqno(II.2)
$$
from a formula of Lienard-Wiechert type in terms of {\it deviation} from
homogeneous distribution of matter along standpoint light cone, {\it nearby}
contributions (on Hubble scale) will dominate.

Before proceeding further let us recall from Reference (1) the quartic
expression for metric {\it at standpoint} ${\bf{R}}$ in terms of $g_{\mu\nu}
(\bf{R})
$,
expressed through the coordinates $x^\mu_{\bf{R}}$ belonging to the
standpoint:
$$
ds^4 = \{[\eta_{\mu\nu} - g_{\gamma\nu} ({\bf{R}}) g^\gamma_\mu ({\bf{R}})]
dx^\mu_{\bf{R}} dx^\nu_{\bf{R}}\}^2 + 4\{g_{\mu\nu}({\bf{R}})
dx^\mu_{\bf{R}} dx^\nu_{\bf{R}}\}^2.\eqno(II.3)
$$
Despite the non-Riemannian character of (II.3), we shall refer to
$g_{\mu\nu}({\bf{R}})$
as ``metric tensor'' because, for small $h_{\mu\nu}({\bf{R}})$
when only terms up to first order in $h_{\mu\nu}(\bf{R})$ are kept,
 (II.3) reduces to
the Riemannian form, $ds^2= 2g_{\mu\nu}({\bf{R}})
dx^\mu_{\bf{R}}dx^\nu_{\bf{R}}$.
The factor 2, irrelevant for local physics, affects ``age of
standpoint''.$^{(2)
}$ Only standpoints for which $|h_{\mu\nu}({\bf{R}})|<< 1$ are covered by our
interim prescription.

The coordinates $x^\mu_{\bf{R}}$ of the compact spacetime belonging to the
${\bf{R}}$ standpoint are limited by the
constraint
$$
0 \leq t_{\bf{R}} \pm r_{\bf{R}} \leq 2 R,\eqno(II.4)
$$where
$$
r_{\bf{R}}\equiv |\vec{x}_{\bf{R}}|,\eqno(II.5)
$$
and $R$ is a positive real parameter that controls standpoint age.
This limitation amounts to the interior of a double light cone.
In its own spacetime the ${\bf{R}}$ standpoint locates at $t_{\bf{R}}= R,
\vec{x}_{\bf{R}} = 0$,
i.e., at the double-cone's center.
The origin of this spacetime - - the point $t_{\bf{R}}=0, \vec{x}_{\bf{R}}
=0$ - -  is interpreted as ``big bang'' and is shared with all other standpoint
spacetimes.
References ((1) and (2) give the general rules for mapping of points within one
standpoint spacetime onto other standpoint spacetimes.

Although the present paper considers only metric tensor {\it at} a standpoint
for the spacetime of that standpoint, such specification (with metric
invariance) when given for {\it all} standpoints provides the complete metric
of
the universe.
Notice that {\it general} coordinate transformations - - hallmark of general
relativity - -  fail to be a feature of standpoint cosmology, being
incompatible
with the constraint (II.4).

For Einstein theory, the retarded Lienard-Wiechert gravitational potential at
${\bf{x}}$ generated by a particle of energy-momentum $p^n_\mu = {\bf{p}}^n$ is
$$
h^n_{\mu\nu} ({\bf{x}}) = G {p^n_\mu p^n_\nu\over {\bf{p}}^n\cdot
({\bf{x}}^n-{\bf{x}})},
\eqno(II.6)
$$
where ${\bf{x}}^n$ locates the intersection of ``source-particle'' trajectory
with
the backward light cone of the point ${\bf{x}}$.
The advanced potential is given by a similar formula.
In the sense of Wheeler and Feynman$^{(4)}$ we require a superposition of
advanced and retarded potentials.
That is, the acceleration experienced by a {\it test particle} at ${\bf{x}}$
combines its gravitational interaction with {\it source particles} in past {\it
and} in future.
If a source-particle trajectory intersects both forward and backward light
cones
of the point ${\bf{x}}$ with no intervening change of energy-momentum, its
advanced and retarded potentials at ${\bf{x}}$ are equal; but generally the
retarded contributions from backward cone are independent of advanced
contributions from forward cone.
In Section IV below we comment further on the advanced potential.

Meaning for the term ``particle'' in this paper is classical, merely implying
localization of (positive) energy within some region of 3-space.
We require particle ``size'' to be small compared to distance from standpoint
but otherwise put no upper limit on ``particle diameter'' and make no
requirement as to ``structure''.
Discreteness of the ``particle'' concept is a convenient conceptual
and notational device for representing
``inhomogeneity'' in matter distribution.

Our proposed analogue of (II.6) for standpoint spacetime (and standpoint
Lorentz
frame$^{(1)}$) dovetails with the structure of standpoint cosmology when the
point ${\bf{x}}$ associates with ${\bf{R}}$ standpoint and the 4-vectors
${\bf{p}}^n$
and ${\bf{x}}^n$ associate with particles whose trajectories intersect
the ${\bf{R}}$ standpoint (forward-backward) light cone.
The Lorentz invariant denominator of (II.6) is exactly the quantity
$\Phi_n({\bf{R}})$ that has been called ``action'' in Reference (1).
We propose as approximate gravitational potential at standpoint, generated by a
(light-cone-intersecting) particle at distance from standpoint {\it small} on
Hubble scale (small compared to $R$), the expression
$$
G {p^n_\mu({\bf{R}})p^n_\nu({\bf{R}})\over \Phi_n({\bf{R}})}.\eqno(II.7)
$$
What about particle sources near the horizon - - at distances approaching $R/2$
on backward cone?
Their number is so great according to Reference (1) as to generate a
meaninglessly-huge gravitational potential $h_{00} ({\bf{R}})$
at standpoint if Formula (II.7) is used.
But because ``homogeneous distribution of sources'',
a notion made precise below, corresponds to $h_{\mu\nu} ({\bf{R}}) =0$, and
because
mean energy density ($\sim {1\over GR^2}^{(1)}$) is negligible at galactic and
smaller scales, it is consistent to suppose potential at standpoint to be a sum
over discrete contributions of the form (II.7) {\it minus } a corresponding
continuous contribution from homogeneously-distributed matter.
Such subtraction suppresses the effect on standpoint gravitational potential
from
matter at Hubble-scale distances, while leaving undamped  ``nearby'' matter
sources.

According to Peebles,$^{(5)}$ homogeneity becomes a good approximation in the
present universe at scales greater than $\sim 1\%$ of Hubble scale.
Roughly speaking, then, our proposal calculates gravitational potential
according to (II.7) from particle sources at distances less than $\sim 1 \%$ of
Hubble length and neglects the contribution from sources at larger distances.
Precise meaning for ``homogeneous'' translates the foregoing rough statement
into
a quantitative prescription.
Because angular isotropy is an aspect of ``homogeneity'', subtraction is needed
only for $h_{{00}} ({\bf{R}})$.
For other components of the gravitational potential, large-distance damping of
the discrete sum is automatic.

\newpage

\noindent{\bf III. Homogeneous Distribution of Matter}

\medskip

We here recall the meaning of ``homogeneous'' as presented in References (1)
and
(2) for matter on the standpoint's backward light cone.
Analogous meaning applies to the forward cone.

Spatial location $\vec{r}_{\bf{R}}$ on backward cone, with respect to
standpoint, is related to a convenient dimensionless 3-vector
 $\vec{b}_{\bf{R}}, 0\leq | \vec{b}_R| \leq 1$,
by the formula
$$
\vec{r}_{\bf{R}} = {1 + \sqrt{2} b_{\bf{R}}\over (1+ b_{\bf{R}})^2}
\tau_{\bf{R}}
\vec{b}_{\bf{R}} ,\eqno(III.1)
$$
where the quantity
$$
\tau_{\bf{R}} = \Omega^{1/2} R,\eqno(III.2)
$$
with
$$
\Omega\equiv ({1\over \sqrt{2}} + \half)^{-2},\eqno(III.3)
$$
is age of standpoint.
The dimensionless factor $\Omega$ gives mean energy density at standpoint
(in the spacetime of that standpoint)  in
units of the standard model's ``critical density''.
That is,$^{(1)}$
$$
\eqalignno{
\rho_{\bf{R}}&= {3\over 8\pi} {1\over GR^2}\cr
           &= \Omega {3\over 8\pi} {H^2_0\over G},&(III.4)\cr}
$$
with
$$
H_0 = \tau^{-1}_{\bf{R}}.\eqno(III.5)
$$
The parameter $\vec{b}_{\bf{R}}$ is useful, partly because it
 controls (through a formula given in Reference (1)) the ``Hubble-flow''
velocity of homogeneously-distributed matter, but especially
because expression through $\vec{b}_{\bf{R}}$ of ``homogeneous distribution'
is simple, as is red shift observed at standpoint for light emitted by
Hubble-flowing
matter.
(It has been shown$^{(1)}$ that $\vec{b}_{\bf{R}}$ is Hubble-flow velocity of
source
in a ``flat'' spacetime belonging to a
standpoint of infinite age but the same spatial location as the ${\bf{R}}$
standpoint.)

Defining a boost (or rapidity) parameter $0\leq \beta_{\bf{R}} \leq \infty$,
equivalent to $b_{\bf{R}}$, by
$$
\beta_{\bf{R}} \equiv \ln \left({1+ b_{\bf{R}}\over 1-b_{\bf{R}}
}\right)^{1/2},\eqno(III.6)
$$
redshift $z_{\bf{R}}$ is given by$^{(2)}$
$$
z_{\bf{R}}= e^{\beta_{\bf{R}}} -1,\eqno(III.7)
$$
while ``homogeneous distribution of matter'' in boost space is proportional to
$$
sinh^2 \beta_{\bf{R}} d\beta_{\bf{R}}.\eqno(III.8)
$$
Notice that, for distance of matter from standpoint small compared to $R$,
$$
\vec{r}_{\bf{R}} \approx \tau_{\bf{R}} \vec{b}_{\bf{R}},\eqno(III.9)
$$
and
$$
b_{\bf{R}} \approx \beta_{\bf{R}} \approx z_{\bf{R}},
$$
while, at the other extreme, ``horizon'' corresponds to
$$
\eqalignno{
r_{\bf{R}} &\to R/2,\cr
b_{\bf{R}} &\to 1,\cr
\beta_{\bf{R}}&\to \infty.&(III.10)
\cr}
$$
An indefinite amount of matter in the universe is seen from (III.8)
 to concentrate  near horizon.
(A useful idea is that age of matter on standpoint backward light
cone is $e^{-\beta_{\bf{R}}}\tau_{\bf{R}}$; ``Mach sources'' on the
standpoint backward cone are thus extremely ``young''.)

Using (III.4) for normalization, one easily computes the increment of
 potential $\Delta (b_{max})$ to be subtracted when calculating $h_{00} (R)$
through (II.7),
corresponding to ``nearby'' homogeneously-distributed matter at distance {\it
less}
than $r_{max}$,
$$
\eqalignno{
\Delta (b_{max})&\approx - {3\over 4} \Omega b^2_{max}\cr
{\hbox{ for}} \ r_{max} &<< R\ (b_{max} << 1).&(III.11)\cr}
$$
As an example, for distances of the order of earth-sun separation
 such that $b_{max} \sim 10^{-14}$, the increment to be subtracted
has order of magnitude $\Delta (b_{max}) \sim 10^{-28}$.
Such a correction may be compared to the much larger, although still small,
gravitational potential at earth generated by sun - - of order $10^{-8}$.
The homogeneous-matter subtraction at earth standpoint continues to be
negligible for all galactic sources, although growing in relative importance
with increasing scale.
Only as $b_{max}$ approaches $10^{-2}$, does the subtraction begin
substantially
to damp the gravitational potential given by (II.7).

\newpage

\noindent{\bf IV. Discussion}

\medskip

The interim prescription proposed here for weak local gravity
facilitates inhomogeneity phenomenology in standpoint cosmology.
With respect to galaxy and star formation, our prescription sustains
previous work based on standard Einstein theory.
Only for scales approaching Hubble scale is there novelty.

With regard to our prescription's use of {\it advanced}
as well as retarded potential, it may be recalled that a similar
electromagnetic
proposal by Wheeler and Feynman$^{(4)}$ implied a boundary condition of
{\it all} ``radiation'' from any particle being eventually absorbed by
a ``sink'' of  other particles.
Such a notion dovetails in standpoint cosmology with indefinite accumulation
near ``future abyss'' of matter serving as ``gravitational sink''.$^{(1,2)}$
As explained in References (1) and (2), ``abyss'' limits the standpoint's
forward
light cone in a sense paralleling that by which ``horizon'' limits the
backward.

Not addressed in this paper are inhomogeneities so large that associated
gravitational
potential $h_{00} ({\bf R})$ approaches or exceeds magnitude 1 - -
inhomgeneities that
in standard parlance are called ``black holes''.
Near such ``strong'' inhomogeneities meaning of ``standpoint light cone''
becomes obscure.
We are hopeful of eventual illumination,
through the non-Riemannian character of (II.3), of
``black-hole'' mysteries arising in standard theory.

Because, in the quantum underpinning of standpoint cosmology, meaning for
spacetime attaches to particle location with respect to
standpoint and meaning for localized energy-momentum attaches to ``particle'',
it is expected that summing (II.7), or some variant thereof, over all particles
 encompasses
the full source of metric at standpoint. Meaning for ``particle
momentum-energy'', that is, includes local ``gravitational
potential.''
(It is not generally true that $\vec{p}^n/E^n$ is particle velocity.)
There is no ``source of metric'' apart from particles.

Helpful discussions with J. Finkelstein, J. D. Jackson, S. Mandelstam,  and
H.P. Stapp are
acknowledged.
This work was supported by the Director, Office of Energy
Research, Office of High Energy and Nuclear Physics, Division of High
Energy Physics of the U.S. Department of Energy under Contract
DE-AC03-76SF00098.

\newpage

\noindent{References}

\begin{enumerate}
\item G. F. Chew , ``Standpoint Cosmology'', Lawrence Berkeley Laboratory
preprint LBL-36314, December, 1994, submitted to Physical Review D.
\item G. F. Chew, ``Predictions from a Single-Parameter Alternative
Cosmology'',
Lawrence Berkeley Laboratory preprint LBL-36554, December, 1994, submitted to
Physical Review Letters.
\item S. Weinberg, ``Gravitation and Cosmology'', Wiley, N.Y., 1972.
\item J. Wheeler and R.P. Feynman, Rev. Mod. Phys. {\bf 21}, 425 (1949).
\item P.J.E. Peebles, ``Principles of Physical Cosmology'', Princeton Univ.
Press, 1993.
\end{enumerate}
\end{document}